\documentclass[preprint,aps,prd,floatfix,nofootinbib]{revtex4}
\usepackage{graphicx}
\usepackage{epsfig}
\usepackage{amsmath}
\usepackage{amssymb}
\usepackage{bm}
\usepackage{bbm}
\begin{document}
\title{\mbox{}\\[10pt]
NRQCD matrix elements for $\bm{S}$-wave bottomonia and
\\
$\bm{\Gamma[\eta_b(nS)\to\gamma\gamma]}$
with relativistic corrections
}
\author{Hee~Sok~Chung}
\affiliation{Department of Physics, Korea University, Seoul 136-701, 
Republic of Korea}
\author{Jungil~Lee}
\affiliation{Department of Physics, Korea University, Seoul 136-701, 
Republic of Korea}
\author{Chaehyun~Yu}
\affiliation{School of Physics, Korea Institute for Advanced Study,
  Seoul 130-722, Republic of Korea}

\begin{abstract}
We determine the leading-order nonrelativistic quantum chromodynamics
(NRQCD) matrix element $\langle \mathcal{O}_1 \rangle_\Upsilon$
and the ratio $\langle \bm{q}^2 \rangle_\Upsilon$, for $\Upsilon=\Upsilon(nS)$
with $n=1$, 2, and 3 by comparing the measured values for 
$\Gamma[\Upsilon\to e^+e^-]$ with the NRQCD factorization formula in which
relativistic corrections are resummed to all orders in the heavy-quark velocity $v$.
The values for $\langle \bm{q}^2 \rangle_{\Upsilon}$, which is the ratio of
order-$v^2$ matrix element to $\langle \mathcal{O}_1 \rangle_\Upsilon$,
are new. They can be used for NRQCD predictions involving $\Upsilon(nS)$ and
$\eta_b(nS)$ with relativistic corrections. As an application, we predict the
two-photon decay rates for the spin-singlet states:
$\Gamma[\eta_b(1S)\to \gamma\gamma] = 0.512^{+0.096}_{-0.094}$ keV,
$\Gamma[\eta_b(2S)\to \gamma\gamma] = 0.235^{+0.043}_{-0.043}$ keV, and
$\Gamma[\eta_b(3S)\to \gamma\gamma] = 0.170^{+0.031}_{-0.031}$ keV.
\end{abstract}

\maketitle


\section{Introduction 
\label{sec:intro}}
The pseudoscalar bottomonium $\eta_b(1S)$, which is the spin-singlet
$S$-wave ground state, was first observed in the photon energy spectrum
of the radiative $\Upsilon(3S)$ decay \cite{:2008vj} and confirmed in
the radiative $\Upsilon(2S)$ decay \cite{:2009pz} by the BABAR Collaboration.
The state was also confirmed by the CLEO Collaboration again in 
$\Upsilon(3S)\to\gamma\eta_b(1S)$ \cite{Bonvicini:2009hs}. So far, only the
mass for the $\eta_b(1S)$ is known as
$m_{\eta_b(1S)} = 9390.9\pm 2.8~\textrm{MeV}$ \cite{Nakamura:2010zzi},
and any of its exclusive decay modes has not been observed, yet.
Among its various decay modes, recent theoretical studies have been
concentrated on relatively clean channels like
$\eta_b \to J/\psi\, J/\psi$~\cite{Jia:2006rx,%
Gong:2008ue,Braguta:2009xu,Santorelli:2007xg}, 
$\eta_b\to J/\psi\,\gamma$~\cite{Hao:2006nf},
and others \cite{Jia:2009ip,Azevedo:2009dt}.\footnote{
Throughout this Letter we use collective notations $\eta_b$ and $\Upsilon$
that indicate $\eta_b(nS)$ and $\Upsilon(nS)$, respectively,
for $n=1,$ 2, and 3 unless a specific state is specified.
} On the other hand,
the most elementary exclusive decay channel is $\eta_b\to\gamma\gamma$,
although it has a large background.
With the decay mode $\Upsilon\to e^+e^-$ of the spin-triplet partner,
$\eta_b\to\gamma\gamma$ must be well described by the nonrelativistic
quantum chromodynamics (NRQCD) factorization formulas for the
electromagnetic decay of heavy quarkonia \cite{Bodwin:1994jh}.
If one makes use of the heavy-quark spin symmetry, then one can make a rough
estimate of the decay rate, whose branching fraction is $\sim 10^{-5}$,
which is relatively greater than other channels listed above.

Available predictions for the decay rate 
$\Gamma[\eta_b\to\gamma\gamma]$ are based on the potential
model \cite{Godfrey:1985xj,Ackleh:1991dy,Ahmady:1994qf,Schuler:1997yw,%
Fabiano:2002nc,Ebert:2003mu,Lakhina:2006vg},
the Salpeter method~\cite{Kim:2004rz,Giannuzzi:2008pv,Laverty:2009xc},
or the heavy-quark spin symmetry~\cite{Lansberg:2006sy}.
Some of them include the effects of the relativistic corrections and
binding effects and most of the predictions rely on the heavy-quark
spin symmetry between the spin-singlet and -triplet states.
One can estimate the spin dependence of the rate systematically by
making use of the potential NRQCD \cite{Pineda:1997bj,Brambilla:2004jw}:
In Ref.~\cite{Penin:2004ay}, the decay rate was computed to the 
next-to-next-to-leading logarithmic accuracies as
$\Gamma[\eta_b(1S)\to \gamma\gamma] =%
0.659\pm 0.089\,(\textrm{th.})^{+0.019}_{-0.018}\,(\delta\alpha_s)\pm 0.015%
\,(\textrm{exp.})$ keV.
Recently, an updated potential-NRQCD prediction for the decay rate
became available: $ 0.54\pm 0.15$ keV \cite{Kiyo:2010zz},
in which leading relativistic corrections are included.

In the mean time, there has been a significant progress in the NRQCD
calculations for $S$-wave charmonium production and decay, in which
relativistic corrections of all orders in the heavy-quark velocity $v$
are resummed \cite{Bodwin:2006dn}. Precise determination of the wavefunction
at the origin for the $J/\psi$ was made based on this 
method \cite{Chung:2007ke,Bodwin:2007fz,etac,Bodwin:2008vp}.
This method has been applied to reconcile the large discrepancy between
the theoretical prediction and the experimental results 
for the cross section $\sigma[e^+e^-\to J/\psi+\eta_c]$ at
the $B$ factories \cite{Bodwin:2006ke,Bodwin:2007ga,He:2007te}.
Therefore, it is worthwhile to improve the NRQCD prediction for
$\Gamma[\eta_b\to\gamma\gamma]$ by taking into account the
relativistic corrections to all orders in $v$. In order to carry out
such an analysis, one needs to know the values for the color-singlet
NRQCD matrix elements and those involving relativistic corrections.
The NRQCD matrix element for the $S$-wave bottomonia can be 
determined by making use of the measured values for
$\Gamma[\Upsilon\to e^+ e^-]$ up to corrections of spin-symmetry breaking
effects. Unfortunately, available order-$v^2$ NRQCD matrix element that
has been fixed from lattice QCD simulations \cite{Bodwin:1996tg,Bodwin:2001mk}
suffers from large uncertainties originated from slow convergence
of the cut-off regularization method.

In this Letter, we first determine the NRQCD matrix elements for
the $S$-wave bottomonium states that are required to compute the
relativistic corrections with considerably less uncertainties
than available values, extending the method in
Refs.~\cite{Bodwin:2006dn,Bodwin:2007fz}.
As an application, we compute $\Gamma[\eta_b\to\gamma\gamma]$,
in which corrections of order the strong coupling $\alpha_s$ and
relativistic corrections of all orders in $\alpha_s^0\bm{q}^{2n}$
are included. Here, $\bm{q}$ is half the relative momentum of $b$
and $\bar{b}$ in the bottomonium rest frame.
The remainder of this Letter is organized as follows:
In Section~\ref{sec:ME}, we estimate the NRQCD matrix elements for the
$S$-wave bottomonium states by making use of the resummed
NRQCD factorization formula against empirical data for the spin-triplet
states. Our prediction for $\Gamma[\eta_b\to\gamma\gamma]$ 
is presented in Section~\ref{sec:twophoton} with
the comparison with available predictions and we summarize in
Section~\ref{sec:summary}.
\section{NRQCD matrix elements for the $\bm{\Upsilon(nS)}$
\label{sec:ME}}
In this section, we briefly review the method to determine
the NRQCD matrix elements for $\Upsilon$
at leading and subleading order in $\bm{q}^2$
based on the strategy for the charmonium counterpart 
in Ref.~\cite{Bodwin:2007fz}. The results are compared with 
those of lattice QCD calculations.

The NRQCD factorization formula for the electromagnetic decay of
the $S$-wave quarkonium $H$ is a linear combination of nonperturbative
NRQCD matrix elements $\langle\mathcal{O}_n\rangle_H$ that are 
classified in powers of $v$, where $\mathcal{O}_n$ is the NRQCD
operator. The factorization is achieved at the amplitude level and
the ratio $\langle \bm{q}^{2n}\rangle_H$ of the order-$\bm{q}^{2n}$
matrix element to the leading one are all, in general, independent.
In addition, the ratios $\langle \bm{q}^{2n}\rangle_H$ have 
power-ultraviolet divergences that must be regulated and, therefore,
the values can even be negative under subtraction. In lattice QCD
calculations, this subtraction is made by making use of the hard-cut-off
regularization whose convergence is slow, resulting in large
uncertainties \cite{Bodwin:2006dn}. 
However, in an electromagnetic decay, in which the color-singlet
contributions dominate, one can calculate the quarkonium wavefunction
of the leading heavy-quark-antiquark ($Q\bar{Q}$) Fock state up to
corrections of relative order $v^2$ if one knows the static,
spin-independent $Q\bar{Q}$ potential exactly. The authors of
Refs.~\cite{Bodwin:2006dn,Bodwin:2007fz} have constructed 
the generalized version,
$\langle \bm{q}^{2n}\rangle_H=[\langle \bm{q}^{2}\rangle_H]^n$,
of the Gremm-Kapustin relation \cite{Gremm:1997dq} to resum a
class of relativistic corrections. The method has been devised to
be consistent with dimensional regularization of these power-ultraviolet
divergent matrix elements.

The resultant formula for the decay rate of $\Upsilon\to e^+e^-$,
in which relativistic corrections of all color-singlet
$Q\bar{Q}$ operator matrix elements are resummed, is given 
by~\cite{Chung:2007ke,Bodwin:2007fz}
\begin{equation}
\Gamma[\Upsilon\to e^+ e^-] =
\frac{ 8 \pi \alpha^2}{27 m_\Upsilon^2}
\left[ 1 - f( \langle v^2 \rangle_\Upsilon  )
- \frac{8\alpha_s}{3\pi} \right]^2 
\langle \mathcal{O}_1 \rangle_\Upsilon,
\label{leptonic}%
\end{equation}
where $m_\Upsilon$ is the $\Upsilon$ mass,
$\langle \mathcal{O}_1 \rangle_\Upsilon$ is the color-singlet 
NRQCD matrix element for the electromagnetic decay of the $\Upsilon$
at leading order in $v$, and 
$\langle v^2\rangle_\Upsilon\equiv\langle\bm{q}^2\rangle_\Upsilon/m_b^2$
with the bottom-quark mass $m_b$. The resummed relativistic corrections
to all orders in $v$ at order $\alpha^2\alpha_s^0$ are contained in
the function $f(x)=x/[3(1+x+\sqrt{1+x})]$ with
$x=\langle v^2 \rangle_\Upsilon$ and in the factor $1/m_\Upsilon^2$
implicitly. 

The order-$\alpha_s^2$ corrections to $\Gamma[\Upsilon\to e^+ e^-]$ 
(Refs.~\cite{Czarnecki:1997vz, Beneke:1997jm}) contain a
strong dependence on the NRQCD factorization scale. If one were to include
those corrections in Eq.~(\ref{leptonic}) and use it to determine $\langle
\mathcal{O}_1 \rangle_\Upsilon$, then $\langle \mathcal{O}_1 \rangle_\Upsilon$ 
would also contain a strong dependence on the NRQCD factorization scale, 
which would cancel in other quarkonium decay and production processes only if
the short-distance coefficients were calculated through relative order
$\alpha_s^2$. Generally, short-distance coefficients for quarkonium processes 
have not been calculated beyond relative order $\alpha_s$. 
For this reason, we omit the order-$\alpha_s^2$ corrections to the 
leptonic width in Eq.~(\ref{leptonic}).
Nevertheless, if one includes the order-$\alpha_s^2$ corrections
and take the factorization scale to be $m_b$, the resultant NRQCD matrix 
elements are increased by about a factor of 40\%.

We briefly discuss the method employed in this Letter to compute 
$\langle \mathcal{O}_1 \rangle_\Upsilon$ and 
$\langle \bm{q}^2 \rangle_\Upsilon$. 
We follow the method given in Ref.~\cite{Bodwin:2007fz} 
and make use of the Cornell potential model~\cite{Eichten:1978tg}. 
By using the Schr\"odinger equation we can express $\langle \mathcal{O}_1
\rangle_\Upsilon$ and $\langle \bm{q}^2 \rangle_\Upsilon$
as functions of the parameters of the Cornell potential model, which are 
the mass parameter in the Schr\"odinger equation, 
the string tension, and the Coulomb strength of the Cornell potential.
The mass parameter can be expressed in terms of the $1S$-$2S$ mass
splitting~\cite{Bodwin:2007fz},
which we compute from the masses of $\Upsilon(1S)$ and $\Upsilon(2S)$. 
The value of the string tension, which is universal, 
is taken from lattice measurements
as $0.1682 \pm 0.0053$ GeV$^2$~\cite{Booth:1992bm, Gupta:1996sa,%
Kim:1993gc, Kim:1996cz}.
Finally, the Coulomb strength parameter is determined by constraining 
the rate (\ref{leptonic}) to be consistent with the experimental 
value~\cite{Nakamura:2010zzi}
and solving the resulting nonlinear equation numerically. 
Because of this, the value of the Coulomb strength parameter is chosen
differently for each quarkonium.
From the fixed values of the model parameters we obtain the numerical values of
the matrix elements.
For details of the method, we refer the readers to Ref.~\cite{Bodwin:2007fz}
and references therein.

We list the numerical values and uncertainties of the parameters used in 
Eq.~(\ref{leptonic}).
The measured leptonic widths of $\Upsilon(nS)$ are
$\Gamma[\Upsilon(1S) \to e^+ e^-] = 1.340 \pm 0.018$\,keV,
$\Gamma[\Upsilon(2S) \to e^+ e^-] = 0.612 \pm 0.011$\,keV, and
$\Gamma[\Upsilon(3S) \to e^+ e^-] = 0.443 \pm 0.008$
\,keV~\cite{Nakamura:2010zzi}.
The masses for the $\Upsilon(nS)$ states are taken to be
$m_{\Upsilon(1S)}=9.46030$ GeV, 
$m_{\Upsilon(2S)}=10.02326$ GeV, and 
$m_{\Upsilon(3S)}=10.3552$ GeV~\cite{Nakamura:2010zzi}, where the
errors ($\lesssim 5\times 10^{-3}\,\%$) are neglected. The factorization
formula (\ref{leptonic}) depends on $m_b$ implicitly through 
$\langle v^2\rangle_\Upsilon$,
where we use the one-loop pole mass $m_b=4.6\pm 0.1$ GeV.
We evaluate $\alpha(\mu)$ and $\alpha_s(\mu)$ at the scale, the momentum
transfer at the quarkonium-photon vertex. 
The values are $\alpha(\mu)=1/131$ in every case,
$\alpha_s[m_{\Upsilon(1S)}] = 0.180\pm 0.032$,
$\alpha_s[m_{\Upsilon(2S)}]=0.177\pm 0.031$, and 
$\alpha_s[m_{\Upsilon(3S)}]=0.176\pm 0.031$, where
the uncertainties of relative order $\alpha_s$ are included in the
strong coupling. 
The main difference between this analysis and that
for the $S$-wave charmonium in Ref.~\cite{Bodwin:2007fz} is that there
are no measured data for $\Gamma[\eta_b\to \gamma\gamma]$.
Therefore, we use the spin-triplet data only.  

By carrying out these calculations, 
the Coulomb strength parameter is fixed as
$9.955$ for $\Upsilon(1S)$, 
$10.960$ for $\Upsilon(2S)$, and
$11.127$ for $\Upsilon(3S)$, respectively.
From these 
we obtain our results for 
$\langle \mathcal{O}_1 \rangle_\Upsilon$ and 
$\langle \bm{q}^2 \rangle_\Upsilon$, which are tabulated in
Table~\ref{table1}.
The corresponding values for the quantity
$\langle v^2 \rangle_\Upsilon $ are 
$\langle v^2 \rangle_{\Upsilon(1S)} = -0.009^{+0.003}_{-0.003}$,
$\langle v^2 \rangle_{\Upsilon(2S)} = 0.090^{+0.011}_{-0.011}$, and
$\langle v^2 \rangle_{\Upsilon(3S)} = 0.155^{+0.018}_{-0.018}$.
These values are in rough agreements with the typical estimate
$v^2 \sim 0.1$ for the bottomonium except that
$\langle v^2 \rangle_{\Upsilon(1S)}$ is tiny.
The error bars in Table~\ref{table1} reflect the uncertainties
arising from $m_b$, $\Gamma[\Upsilon\to e^+e^-]$, 
string tension, $\alpha_s$, and the ignorance
of the spin-dependent interactions of the potential in the 
Schr\"odinger equation~\cite{Bodwin:2007fz}, all of which are
added in quadrature.
The values for the leading-order matrix elements 
$\langle \mathcal{O}_1 \rangle_\Upsilon$
in Table~\ref{table1} have been used to predict the inclusive
charm production in $\Upsilon$ decays \cite{Kang:2007uv}
and the heavy quarkonium production associated with a photon 
in $e^+ e^-$ annihilation~\cite{Chung:2008km}.
The values for $\langle \bm{q}^2 \rangle_\Upsilon$ in 
Table~\ref{table1} are new.
The value for $\Upsilon(1S)$ has errors 
significantly less than those of the available lattice QCD
calculations \cite{Bodwin:1996tg,Bodwin:2001mk}.
One can reduce theoretical uncertainties by considering
the dependence on $m_b$ that also appears
in the short-distance coefficients of factorization formulas.
Therefore, we present the sources of errors in Table~\ref{table1}.
Unlike the $S$-wave charmonium case in Ref.~\cite{Bodwin:2007fz},
the uncertainties of $\langle \mathcal{O}_1 \rangle_\Upsilon$ and
$\langle \bm{q}^2 \rangle_\Upsilon$ due to the errors of the
heavy-quark mass are insignificant.
\begin{table}[t]
\caption{
\label{table1}%
The NRQCD matrix element $\langle \mathcal{O}_1 \rangle_{\Upsilon}$
at the leading order in $v$ in units of GeV$^3$
and ratios $\langle \bm{q}^2 \rangle_\Upsilon$ in units of GeV$^2$
for $\Upsilon=\Upsilon(1S)$, $\Upsilon(2S)$, and $\Upsilon(3S)$.
}
\begin{ruledtabular}
\begin{tabular}{lc|cccccc}
Sources of errors&
&
$\langle \mathcal{O}_1 \rangle_{\Upsilon(1S)}$&
$\langle \bm{q}^2      \rangle_{\Upsilon(1S)}$&
$\langle \mathcal{O}_1 \rangle_{\Upsilon(2S)}$&
$\langle \bm{q}^2      \rangle_{\Upsilon(2S)}$&
$\langle \mathcal{O}_1 \rangle_{\Upsilon(3S)}$&
$\langle \bm{q}^2      \rangle_{\Upsilon(3S)}$
\\
\hline
$\Delta m_b$ &
&
$3.069^{+0.000}_{-0.001}$ &
$-0.193^{+0.000}_{-0.000}$ &
$1.623^{+0.002}_{-0.002}$ &
$1.898^{+0.001}_{-0.000}$ &
$1.279^{+0.003}_{-0.003}$ &
$3.283^{+0.003}_{-0.002}$ 
\\
others &
&
$3.069^{+0.207}_{-0.190}$ &
$-0.193^{+0.072}_{-0.073}$ &
$1.623^{+0.112}_{-0.103}$ &
$1.898^{+0.210}_{-0.210}$ &
$1.279^{+0.090}_{-0.083}$ &
$3.283^{+0.353}_{-0.352}$ 
\\
\hline
total &
&
$3.069^{+0.207}_{-0.190}$ &
$-0.193^{+0.072}_{-0.073}$ &
$1.623^{+0.112}_{-0.103}$ &
$1.898^{+0.210}_{-0.210}$ &
$1.279^{+0.090}_{-0.083}$ &
$3.283^{+0.353}_{-0.352}$ 
\end{tabular}
\end{ruledtabular}
\end{table}

Our results are now compared with those for the ground-state 
$S$-wave bottomonium obtained from a lattice QCD simulation.
The results from the quenched approximation are given in
Ref.~\cite{Bodwin:1996tg}. We quote the updated results
of the unquenched analysis in Ref.~\cite{Bodwin:2001mk}:
The leading-order NRQCD matrix element is
$\langle \mathcal{O}_1 \rangle_{1S}
= 4.10 (1)(9)(41)$ GeV$^3$ and the ratio 
$\langle \bm{q}^2 \rangle_{1S}$
ranges from about $-5$\,$\textrm{GeV}^2$ to about 
2$\,\textrm{GeV}^2$ \cite{Bodwin:2001mk}.
Here, the subscript $1S$ indicates the average of $\eta_b(1S)$ and $\Upsilon(1S)$.
Our central value for the $\langle \mathcal{O}_1 \rangle_{\Upsilon(1S)}$ 
is about $25\,$\% smaller than that of Ref.~\cite{Bodwin:2001mk},
which is greater than the quenched case \cite{Bodwin:1996tg} 
by about a factor of 2. In the case of
$\langle \bm{q}^2 \rangle_{\Upsilon(1S)}$,
our result is consistent with that in Ref.~\cite{Bodwin:2001mk}
but ours has uncertainties significantly smaller than that of
the lattice result.
\section{Two-photon widths for the $\bm{\eta_b}$ 
\label{sec:twophoton}}
In this section, we predict $\Gamma[\eta_b \to \gamma\gamma]$
by making use of the NRQCD matrix elements determined in Section~\ref{sec:ME}.
In fact, the NRQCD matrix element 
$\langle \mathcal{O}_1 \rangle_{\eta_b}$ that appears in the
factorization formula for $\Gamma[\eta_b \to \gamma\gamma]$ might
be different from $\langle \mathcal{O}_1 \rangle_{\Upsilon}$ by
a relative order $v^2$, which breaks the approximate heavy-quark spin 
symmetry \cite{Bodwin:1994jh}. We recall that the 
effect of spin-symmetry breaking in the low-lying $S$-wave charmonia
$J/\psi$ and $\eta_c$ is not significant~\cite{Bodwin:2007fz}.
Therefore, the errors in the approximation 
$\langle \mathcal{O}_1 \rangle_{\eta_b(nS)}\approx%
\langle \mathcal{O}_1 \rangle_{\Upsilon(nS)}$ may be insignificant
based on the fact that $\langle v^2 \rangle_\Upsilon \ll%
\langle v^2 \rangle_{J/\psi}=0.22$~\cite{Bodwin:2007fz}.

\begin{table}[t]
\caption{
\label{table2}%
Two-photon widths of the $\eta_b(nS)$ in units of keV.
}
\begin{ruledtabular}
\begin{tabular}{l|ccc}
State & 
$\eta_b(1S)$&
$\eta_b(2S)$&
$\eta_b(3S)$
\\
\hline
$\Gamma_{\gamma\gamma}$ 
& $0.512^{+0.096}_{-0.094} $ 
& $0.235^{+0.043}_{-0.043} $
& $0.170^{+0.031}_{-0.031} $
\\
\end{tabular}
\end{ruledtabular}
\end{table}
As in the leptonic decay of the $\Upsilon$, we include the relativistic 
corrections to $\Gamma[\eta_b \to \gamma\gamma]$ to all orders in $v$.
The resultant factorization formula is given by~\cite{etac,Bodwin:2007fz}
\begin{equation}
\Gamma[\eta_b \to \gamma\gamma] =
\frac{2\pi\alpha^2}{81 m_b^2}
\left[
1 - g(\langle v^2 \rangle_{\eta_b} ) 
- \frac{(20-\pi^2)\alpha_s}{6\pi}
\right]^2
\langle \mathcal{O}_1 \rangle_{\eta_b},
\label{twophoton}%
\end{equation}
where the relativistic corrections are incorporated into the function 
$g(x)=1-\{\log[1+2\sqrt{x(1+x)}+2x]\}/[2\sqrt{x(1+x)}]$ with 
$x=\langle v^2\rangle_{\eta_b}\equiv\langle\bm{q}^2\rangle_{\eta_b}/m_b^2$.
The input parameters for the numerical calculations are chosen in a similar
way in Ref.~\cite{Bodwin:2007fz} for 
$\Gamma[\eta_c \to \gamma\gamma]$. The scale $\mu$ for the couplings
$\alpha$ and $\alpha_s$ are taken to be the momentum transfer at the
photon-heavy-quark vertex, namely, $m_{\eta_b}/2$:
$\alpha=1/132$ for every case,
$\alpha_s[m_{\eta_b(1S)}/2]=0.216\pm 0.046$,
$\alpha_s[m_{\eta_b(2S)}/2]=0.212\pm 0.045$, and
$\alpha_s[m_{\eta_b(3S)}/2]=0.210\pm 0.044$,
where the uncertainties of the strong coupling are of relative order $\alpha_s$.
For the meson masses, we use
$m_{\eta_b(1S)} = 9390.9\pm 2.8~\textrm{MeV}$ \cite{Nakamura:2010zzi},
$m_{\eta_b(2S)}=9.97$ GeV, and $m_{\eta_b(3S)}=10.3$ GeV, where we have
assumed that $m_{\Upsilon(nS)}-m_{\eta_b(nS)}=0.5$\,MeV for $n=2$ and $3$.
While this value for the hyperfine mass splitting is smaller than the
measured value for the $1S$ states $m_{\Upsilon(1S)}-m_{\eta_b(1S)}=%
69.3\pm2.8$\,MeV~\cite{Nakamura:2010zzi}, it is comparable to 
that for the $2S$ charmonia,
$m_{\psi(2S)}-m_{\eta_c(2S)}=49$\,MeV~\cite{Nakamura:2010zzi}.
Note that the uncertainties from $m_{\eta_b(2S)}$ and $m_{\eta_b(3S)}$ are 
insignificant because the factorization formula (\ref{twophoton}) does not
depend on them but on $m_b$.
Like the leptonic width of the $\Upsilon$ [Eq.~(\ref{leptonic})], we omit the
order-$\alpha_s^2$ corrections to the two-photon width of the $\eta_b$, whose
result is available in Ref.~\cite{Czarnecki:2001zc}.

The resultant predictions for 
$\Gamma[\eta_b\to\gamma\gamma]$ are tabulated
in Table~\ref{table2}. The errors include the uncertainties
of $\alpha_s$, $m_b$, and the values for
$\langle \mathcal{O}_1 \rangle_{\Upsilon}$ and
$\langle \bm{q}^2 \rangle_{\Upsilon}$ in Table~\ref{table1}.
We also include the errors of using 
$\langle \mathcal{O}_1 \rangle_{\Upsilon}$ and
$\langle \bm{q}^2 \rangle_{\Upsilon}$, which are of relative
order $v^2$ set to be 0.1.
From the order-$\alpha_s^2$ corrections to the electromagnetic widths of the
$\Upsilon$ and $\eta_b$~\cite{Czarnecki:1997vz,Beneke:1997jm,Czarnecki:2001zc}, 
we find that the order-$\alpha_s^2$ corrections 
account for $-2.64\,\alpha_s^2$ in the ratio 
$\Gamma[\eta_b \to \gamma \gamma] / \Gamma[\Upsilon \to e^+ e^-]$, 
if we choose the NRQCD factorization scale to be
$m_b$. Therefore, we include the errors of omitting the order-$\alpha_s^2$
corrections in using $\Gamma[\Upsilon \to e^+ e^-]$ to determine
$\Gamma[\eta_b \to \gamma \gamma]$ as $2.64\,\alpha_s^2$.
This implies that the large correction to the leading-order
NRQCD matrix elements arising from inclusion of the order-$\alpha_s^2$
corrections, as briefly shown in the previous section, almost cancels
the order-$\alpha_s^2$ corrections to the two-photon width of the $\eta_b$.
All of the errors listed above are added in quadrature.
We can compare our results with previous predictions.
In the case of $\eta_b(1S)$, available predictions range 
from $0.170$ keV to $0.659$ keV. The results in 
Refs.~\cite{Ahmady:1994qf,Schuler:1997yw,%
Fabiano:2002nc,Ebert:2003mu,Kim:2004rz,Giannuzzi:2008pv,Laverty:2009xc,%
Lansberg:2006sy} agree with our prediction within errors, while
some models~\cite{Godfrey:1985xj,Ackleh:1991dy,Lakhina:2006vg},
which does not use the heavy-quark spin symmetry,
apparently underestimate the rate in comparison with ours.
Our result $\Gamma[\eta_b(1S)\to\gamma\gamma]=0.512^{+0.096}_{-0.094}$
agrees with the most recent potential-NRQCD prediction
$0.54\pm 0.15$ keV in Ref.~\cite{Kiyo:2010zz} in which the leading relativistic
corrections are included, while it is smaller than another potential-NRQCD
prediction in Ref.~\cite{Penin:2004ay}.
Note that we have borrowed $\langle \mathcal{O}_1\rangle_\Upsilon$ and
$\langle \bm{q}^2\rangle_\Upsilon$ for
$\langle \mathcal{O}_1\rangle_{\eta_b}$ and
$\langle \bm{q}^2\rangle_{\eta_b}$ in Eq.~(\ref{twophoton})
after taking into account the errors
of spin-symmetry breaking effect as $v^2\sim 0.1$
because $\Gamma[\eta_b\to\gamma\gamma]$ are not measured. Once 
$\Gamma[\eta_b(nS)\to \gamma\gamma]$ are measured in the future,
one can determine 
$\langle \mathcal{O}_1 \rangle_{\eta_b(nS)}$ and
$\langle \bm{q}^2 \rangle_{\eta_b(nS)}$ (or 
eventually $\langle v^2 \rangle_{\eta_b(nS)}$) with an improved accuracy
in combination with the measured values for $\Gamma[\Upsilon\to e^+e^-]$.

\section{Summary
\label{sec:summary}}
In summary, we have determined the leading-order
NRQCD matrix element $\langle \mathcal{O}_1 \rangle_\Upsilon$
and the ratio $\langle \bm{q}^2 \rangle_\Upsilon$, for $\Upsilon=\Upsilon(nS)$
with $n=1$, 2, and 3 by comparing the measured values for the leptonic
decay rates of the $\Upsilon$ with the NRQCD factorization formula in which
relativistic corrections to all orders in $v$ are included. The values
for $\langle \bm{q}^2 \rangle_{\Upsilon}$ are new and can be used for
various phenomenological predictions for $\Upsilon$ and $\eta_b$ including
relativistic corrections. The values for
$\langle \bm{q}^2 \rangle_{\Upsilon}$ are consistent with the naive expectation
of the velocity-scaling rules except that 
$\langle \bm{q}^2 \rangle_{\Upsilon(1S)}$ is tiny.
By assuming approximate heavy-quark spin symmetry with the uncertainties
of relative order $v^2\sim 0.1$, we used  
$\langle \mathcal{O}_1 \rangle_\Upsilon$
and $\langle \bm{q}^2 \rangle_\Upsilon$ to estimate 
$\Gamma[\eta_b(1S)\to \gamma\gamma] = 0.512^{+0.096}_{-0.094}$ keV,
$\Gamma[\eta_b(2S)\to \gamma\gamma] = 0.235^{+0.043}_{-0.043}$ keV, and
$\Gamma[\eta_b(3S)\to \gamma\gamma] = 0.170^{+0.031}_{-0.031}$ keV.
Our prediction for $\Gamma[\eta_b(1S)\to \gamma\gamma]$ is consistent with
a recent potential-NRQCD prediction in Ref.~\cite{Kiyo:2010zz}, in which
the leading relativistic corrections are included.

By making use of the ratio $\Gamma[\eta_b(1S)\to\gamma\gamma]/%
\Gamma[\eta_b(1S)\to gg]$, one can make a rough estimate of the branching
fraction for $\eta_b(1S)\to\gamma\gamma$ as $\sim 6.9\times 10^{-5}$.
The BABAR Collaboration reported 
$19200\pm 2000\pm 2100$ $\eta_b(1S)$ events
out of $(109 \pm 1)\times 10^{6}$ $\Upsilon(3S)$ samples~\cite{:2008vj}.
They also obtained $12800\pm 3500^{+3500}_{-3100}$ $\eta_b(1S)$ events
from $(91.6\pm 0.9)\times 10^6$ $\Upsilon(2S)$ samples~\cite{:2009pz}.
These are not sufficient to observe 
the mode $\eta_b(1S)\to\gamma\gamma$.
However, we expect that this channel
can be observed at the superKEKB or superB factory if more data are accumulated.
The CERN Large Hadron Collider is expected to produce about 
$5\times 10^9$ $\eta_b$'s with the integrated luminosity 
$\sim 300$ fb$^{-1}$~\cite{Gong:2008ue}, with which one can probe about
$45000$ events of $\eta_b(1S)\to\gamma\gamma$. We anticipate such a stage
against which our predictions can be tested.

\begin{acknowledgments}
This work was supported by the Basic Science Research
Program through NRF of Korea funded by the MEST under
contracts 2010-0015682 (HSC) and 2010-0028228 (JL) and 2009-0072689 (CY).
\end{acknowledgments}

\end{document}